\documentstyle[preprint,pra,aps]{revtex}
\begin{document}
\draft
\title{\bf
Quantum tunnelling of a complex system: effects of a finite size
and intrinsic structure}
\author{V.V. Flambaum$^{1}$ and V.G. Zelevinsky$^{2}$}
\address{$^{1}$School of Physics, University of New South Wales,
Sydney 2052, Australia\\
$^{2}$National Superconducting Cyclotron Laboratory and\\
Department of Physics and Astronomy, Michigan State University,
East Lansing, MI 48824-1321, USA}
\date{\today}
\maketitle

\tightenlines
\begin{abstract}
A simple model is considered to study the effects of finite size
and internal structure in the tunnelling of bound two-body systems
through a potential barrier. It is demonstrated that these effects
are able to increase the tunnelling probability. Applications may
include nuclear fusion, hydrogen atom and Cooper pair tunnelling.
\end{abstract}
\vspace{1cm} 

\section{Introduction}

The processes using the mechanism of quantum tunnelling are
abundant in different areas of physics. Chemical reactions at low
temperatures, diverse phenomena in solid state physics, nuclear
alpha and cluster decays, fission and fusion processes in
thermonuclear reactors, shortly after the Big Bang and inside
stellar matter proceed through the Coulomb barrier. The
conventional textbook treatment is, as a rule, limited by
considering tunnelling of a point-like object. It is known that
the intrinsic degrees of freedom of a system modelled by a barrier
are important, and one needs to take into account the
probabilistic distribution of barriers \cite{balantekin98} in
order to explain the subbarrier fusion. The friction-like
processes under the barrier hinder the tunnelling
\cite{caldeira8183}. At the same time, interaction of a charged
particle with virtual electromagnetic field (an analog of the Lamb
shift in bound states) increases the tunnelling probability
\cite{FZ99}. In reality, in many cases one has to deal with the
tunnelling of a composite object with its own intrinsic degrees of
freedom.

An important question of how the tunnelling reactions are
influenced by the finite size and the structure of the tunnelling
object remains poorly understood. For example, the deuteron is a
weekly bound complex with a size that is large on nuclear scale.
How does this size influence the fusion reaction
$d+d\rightarrow^4$He deeply under the Coulomb barrier? There are
no experimental laboratory data for the fusion cross section at
low ($\sim$ keV) deuteron energies, since the cross section is too
small. However, there is a discrepancy between the standard theory
and the existing measurements for $dd$-reactions in solids at
energy between 10 and 2 keV \cite{yuki98,altshuler01}. This is an
area of extreme importance for astrophysical and thermonuclear
applications, especially with the perspective of radioactive beam
facilities of new generation. Currently, the nuclear cross
sections for low energies are determined by extrapolation based on
the results for point-like tunnelling particles.

Another example of great interest in relation to various
tunnelling devices is given by electron Cooper pairs in
superconductors. In ``usual" superconductors, the size of the pair
is so large that the electron binding is of minor significance for
such processes, and the Josephson tunnelling matrix element can be
taken as a product of penetrabilities for independent electrons.
In high-temperature superconductors, the size of the pair is much
smaller, and the binding may produce corrections. The tunneling
of a bound electron pair through the potential barrier
created at the ``point contact'' in semiconductors is also
 of interest because this process may be related to the
mysterious 0.7($2 e^2/\hbar$) structure in the measurements of
conductance quantization in one-dimensional systems \cite{FK}
(note that a bound pair appears  in a one-dimensional system even
for an arbitrary weak attraction). One can also be interested in
finite size effects for hydrogen tunnelling (a system of one heavy
particle and one light particle).

To provide an analytical estimate of the finite size effect we use
a simple model. A system consists of two particles bound by the
oscillator potential. An additional parabolic potential of
positive or negative sign acts on one of the particles. This model
allows one to study the limiting cases and shows that the finite
size of a tunnelling system may be very important. Similar results
for a pair of particles bound in an infinite potential box, while
one of the particles encounters a rectangular barrier, were found
in the old paper by Zakhariev and Sokolov \cite{zakhariev64}, with
the conclusion that the effective barrier is modified which
facilitates the penetration. We show also that the electrostatic
polarizability of a composite particle adiabatically approaching
the Coulomb barrier increases the tunnelling probability.

\section{Two oscillators}

\subsection{Normal modes}

We consider two particles interacting via the harmonic potential
and moving in the external field that acts on one particle only.
The Hamiltonian of such a system is
\begin{equation}
H=\frac{p_{x}^{2}}{2m_{x}}+\frac{p_{y}^{2}}{2m_{y}}+\frac{1}{2}
k(x-y)^{2}+\frac{1}{2}qx^{2}.               \label{1.1}
\end{equation}
Here coordinates, $x$ and $y$, and corresponding momenta, $p_{x}$
and $p_{y}$, refer to a ``proton'' and ``neutron'', respectively;
the potential $(1/2)k(x-y)^{2}$ describes the binding of the
particles, and the potential $(1/2)qx^{2}$ acts on the ``proton''
only. The case of $q<0$ models a Coulomb potential barrier for the
``proton'' in the process of the fusion of two ``deuterons'' into
$^4$He or fusion of the ``deuteron'' with any other nucleus.

The secular equation for normal modes reads
\begin{equation}
\omega^{4}-\omega^{2}\left(\frac{k}{\mu}+\frac{q}{m_{x}}\right)
+\frac{kq}{m_{x}m_{y}}=0,                   \label{1.6}
\end{equation}
where the reduced mass $\mu=m_{x}m_{y}/(m_{x}+m_{y})$. The
solution may be conveniently written in terms of the parameters of
bare frequencies and the ratio of the masses,
\begin{equation}
\nu^{2}=\frac{k}{\mu}, \quad \Omega^{2}=\frac{q}{m_{x}}, \quad
\lambda=\frac{m_{x}}{m_{x}+m_{y}}.            \label{1.7}
\end{equation}
Then the frequencies of two normal modes are
\begin{equation}
\omega_{\pm}^{2}=\frac{1}{2}\Bigl[\nu^{2}+\Omega^{2}\pm
\sqrt{\nu^{4}+\Omega^{4}+2\nu^{2}\Omega^{2}(1-2\lambda)}\Bigr];
                                              \label{1.8}
\end{equation}
at $q<0$, $\omega_{-}$ is imaginary. The case of equal masses,
$m_{x}=m_{y}=m$, is $\lambda=1/2$, when
\begin{equation}
\omega_{\pm}^{2}=\frac{1}{2}\Bigl[\nu^{2}+\Omega^{2}\pm
\sqrt{\nu^{4}+\Omega^{4}}\Bigr]=\frac{1}{2m}\Bigl[q+2k\pm
\sqrt{q^{2}+4k^{2}}\Bigr].                   \label{1.9}
\end{equation}
If the $x$-particle (``proton") is very light, $m_{x}\rightarrow
0,\,\lambda\rightarrow 0$,
\begin{equation}
\omega_{+}^{2}=\nu^{2}+\Omega^{2}, \quad \omega_{-}^{2}
\rightarrow 0;                              \label{1.10}
\end{equation}
if the $y$-particle (``neutron") is very light, $m_{x}\rightarrow
0,\,\lambda\rightarrow 1$, two bare frequencies are normal modes,
\begin{equation}
\omega_{+}^{2}=\nu^{2}, \quad \omega_{-}^{2}=\Omega^{2}.
                                              \label{1.11}
\end{equation}

\subsection{Wave function}

To find the spatial structure of the normal modes, we carry out
the standard diagonalization of the Hamiltonian. First we make the
scale transformation to new coordinates $u_{x,y}$ and new momenta
$v_{x,y}$,
\begin{equation}
p_{x}=\sqrt{m_{x}}v_{x}, \quad p_{y}=\sqrt{m_{y}}v_{y}, \quad
x=\frac{u_{x}}{\sqrt{m_{x}}}, \quad y=\frac{u_{y}}{\sqrt{m_{y}}}.
                                              \label{2.1}
\end{equation}
Then the Hamiltonian takes the form
\begin{equation}
H=\frac{1}{2}(v_{x}^{2}+v_{y}^{2})+\frac{k}{2}\left(\frac{u_{x}}
{\sqrt{m_{x}}}-\frac{u_{y}}{\sqrt{m_{y}}}\right)^{2}+ \frac{1}{2}
\,\frac{qu_{x}^{2}}{m_{x}}.                      \label{2.2}
\end{equation}
Since the kinetic form is invariant, we diagonalize the potential
form by the orthogonal transformation
\begin{equation}
u_{x}=\xi_{1}\cos\phi+\xi_{2}\sin\phi, \quad
u_{y}=-\xi_{1}\sin\phi+\xi_{2}\cos\phi.         \label{2.3}
\end{equation}
The transformed potential energy is
\begin{equation}
U=\frac{1}{2}\xi_{1}^{2}U_{11}+\frac{1}{2}\xi_{2}^{2}U_{22}+
\xi_{1}\xi_{2}U_{12},                          \label{2.4}
\end{equation}
where
\begin{equation}
U_{11}=k\left(\frac{\cos\phi}{\sqrt{m_{x}}}+\frac{\sin\phi}
{\sqrt{m_{y}}}\right)^{2}+\frac{q}{m_{x}}\cos^{2}\phi, \label{2.5}
\end{equation}
\begin{equation}
U_{22}=k\left(\frac{\sin\phi}{\sqrt{m_{x}}}-\frac{\cos\phi}
{\sqrt{m_{y}}}\right)^{2}+\frac{q}{m_{x}}\sin^{2}\phi, \label{2.6}
\end{equation}
\begin{equation}
U_{12}=k\left(\frac{\cos\phi}{\sqrt{m_{x}}}+\frac{\sin\phi}
{\sqrt{m_{y}}}\right)\left(\frac{\sin\phi}{\sqrt{m_{x}}}-
\frac{\cos\phi}{\sqrt{m_{y}}}\right)+\frac{q}{m_{x}}\cos\phi\,\sin\phi.
                                                  \label{2.7}
\end{equation}
The diagonalization condition $U_{12}=0$ determines the mixing
angle $\phi$,
\begin{equation}
\tan(2\phi)=\frac{2k\sqrt{m_{x}m_{y}}}{qm_{y}+k(m_{y}-m_{x})}=
\frac{2\nu^{2}\sqrt{\lambda(1-\lambda)}}{\Omega^{2}+\nu^{2}(1-2\lambda)}.
                                                   \label{2.8}
\end{equation}
From here
\begin{equation}
\sin(2\phi)=\frac{2\nu^{2}\sqrt{\lambda(1-\lambda)}}{\sqrt{\Omega^{4}
+\nu^{4}+2\nu^{2}\Omega^{2}(1-2\lambda)}},         \label{2.9}
\end{equation}
\begin{equation}
\cos(2\phi)=\frac{\Omega^{2}+\nu^{2}(1-2\lambda)}{\sqrt{\Omega^{4}
+\nu^{4}+2\nu^{2}\Omega^{2}(1-2\lambda)}},         \label{2.10}
\end{equation}
Therefore we get, as it should be, in agreement with (\ref{1.8}),
\begin{equation}
U_{11}=\omega_{+}^{2}, \quad U_{22}=\omega_{-}^{2}. \label{2.11}
\end{equation}
The wave function  of the system is the product of two oscillator
functions,
\begin{equation}
\Psi(x,y)=\psi_{1}(\xi_{1};\omega_{+})\psi_{2}(\xi_{2};\omega_{-}).
                                                   \label{2.12}
\end{equation}

\section{Physical effects}

\subsection{Probability of tunnelling}

The most interesting case corresponds to negative $q$ and, whence,
negative $\omega_{-}^{2}$. The transmission coefficient for the
parabolic barrier is given by \cite{BM2}
\begin{equation}
T=\frac{D}{1+D},                                   \label{3.1}
\end{equation}
where
\begin{equation}
D=\exp\left(\frac{2 \pi E}{\hbar |\omega_{-}|}\right)  \label{3.2}
\end{equation}
is the semiclassical penetrability for a ``one-way" traverse. Here
$E$ is the energy of the ``deuteron'' for motion along the
coordinate $\xi_{2}$, corresponding to the imaginary frequency;
$E=0$ corresponds to the top of the barrier. The tunnelling
probability is exponentially small for $E<0$. A realistic case
corresponds to $k\gg |q|$, a ``deuteron'' bound much stronger than
Coulomb energy. In this case (for equal masses)
\begin{equation}
|\omega_{-}|\approx
\sqrt{\frac{|q|}{2m}}\left(1+\frac{|q|}{8k}\right)\equiv
\omega_{0}\left(1+\frac{|q|}{8k}\right).         \label{3.3}
\end{equation}
The factor $1+|q|/(8k)$ provides an exponential enhancement of the
tunnelling probability for the finite size ``deuteron'' in
comparison with a particle of a zero size ($k=\infty$),
\begin{equation}
D=D_{0}\exp\left(\frac{\pi|E|\sqrt{m|q|}}{2\sqrt{2}\hbar
k}\right)=D_{0}\exp\left(\frac{\pi|E|m\omega_{0}}{2\hbar
k}\right),                                      \label{3.4}
\end{equation}
where
\begin{equation}
D_{0}=\exp\left(-\frac{2\pi|E|}{\hbar\omega_{0}}\right).\label{3.5}
\end{equation}
For the real deuteron this effect is too small but it increases if
the ``charged" particle is very light, $m_{x}/m_{y}\ll 1$,
\begin{equation}
|\omega_{-}|\approx
\sqrt{\frac{|q|}{m_{x}+m_{y}}}\left(1+\frac{|q|}{8k}\,\frac{m_{y}}{m_{x}}
\right).                                        \label{3.6}
\end{equation}

\subsection{Polarizability}

There is another effect that may be even more important than the
modification of the barrier (change of $\omega_{-}$) considered in
the preceding subsection and analogous to the effect of ref.
\cite{zakhariev64}. This is energy transfer from internal motion
of the ``deuteron'' to the translational motion which happens for
a non-parabolic barrier. Let us consider a more realistic model,
where the potential barrier is smoothly goes to zero at infinity
being parabolic near the top. The change of internal motion of the
``deuteron" in the ground state occurs adiabatically as a
transition from the unperturbed value $\hbar \nu/2$ far away from
the barrier to $\hbar \omega_+/2$. This energy difference is
converted into energy of translational motion of the ``deuteron'':
\begin{equation}
E=E(\infty)+\frac{\hbar\nu}{2}-\frac{\hbar\omega_+}{2} \approx
E(\infty)+ \frac{\hbar q}{4\sqrt{km_{x}}}\left(\frac{m_{y}}
{m_{x}+m_{y}}\right)^{3/2}.                    \label{3.7}
\end{equation}
This correction also grows if the ``charged" particle is light.

For the real deuteron (or any charged system) this correction may
be presented as a result of the deuteron polarization,
\begin{equation}
E=E(\infty)+\frac{1}{2}\alpha {\cal E}^2,           \label{3.8}
\end{equation}
where $\alpha$ is the deuteron polarizability and ${\cal E}$ the
electric field produced by the Coulomb field. The internal ground
state energy decreases by this amount, $\delta E=-\alpha {\cal
E}^2 /2$, that is transferred into translational motion. Indeed,
the electric field polarizes the deuteron (the electric dipole
moment is $d=\alpha {\cal E}$). This creates an additional force,
$(d/d x)\alpha{\cal E}^2 /2$. Integration of this force gives a
positive correction to kinetic energy, or, to present it more
conveniently, a negative correction to the Coulomb barrier
potential,
\begin{equation}
\delta U =-\frac{1}{2}\alpha {\cal E}^2 =-\alpha \frac{Z^2 e^2}
{2r^4}.                                             \label{3.9}
\end{equation}

Again, this correction increases the barrier penetration factor.
For the deuteron, the empirical value \cite{rodning82} is
$\alpha$=0.70(5) fm$^3$. We can estimate the change of the Coulomb
barrier penetration factor due to this potential ($U=Ze^2/r
+\delta U$). In the WKB semiclassical approximation,
\begin{equation}
D=\exp\left(-\frac{2\pi Ze^2}{\hbar v}\right)
\exp\left(\frac{\alpha
Z^{3/2}e\sqrt{2m}}{5\hbar(r_c)^{5/2}}\right),        \label{3.10}
\end{equation}
where $v$ is the relative velocity, $m$ the reduced mass for two
nuclei, and $r_c$ the cut-off radius where the validity of the
potential approximated by $U=Ze^2/r +\delta U$ is violated ($r_c
\sim r_d$ where $r_d$ is the deuteron radius). For the fusion of
two deuterons, the correction due to the polarization potential is
small ($\sim$ 1\%). However, it rapidly increases with the nuclear
charge.

\subsection{``Cold fusion'' and muon catalysis}

In a process of muon catalysis, the fusion proceeds from a ground
state of a muonic molecule, $dd\mu$ or $dt\mu$. This process
occurs due to zero-point oscillations in the ground state of the
molecule, as soon as a neutron from one nucleus reaches the strong
potential well of another nucleus. In the two-oscillator model
with $q>0$, the expectation values of the normal coordinates in
the ground states are
\begin{equation}
\langle \xi_{1}^{2}\rangle=\frac{\hbar}{2\omega_{+}}, \quad \langle
\xi_{2}^{2}\rangle=\frac{\hbar}{2\omega_{-}}, \quad \langle
\xi_{1}\xi_{2}\rangle=0.                        \label{3.11}
\end{equation}
Going back to the original coordinates, eqs. (\ref{2.1}) and
(\ref{2.3}), we find
\begin{equation}
\langle x^{2}\rangle=\frac{\hbar}{2m_{x}}\left(\frac{\cos^{2}\phi}
{\omega_{+}}+\frac{\sin^{2}\phi}{\omega_{-}}\right), \label{3.12}
\end{equation}
\begin{equation}
\langle y^{2}\rangle=\frac{\hbar}{2m_{y}}\left(\frac{\cos^{2}\phi}
{\omega_{-}}+\frac{\sin^{2}\phi}{\omega_{+}}\right), \label{3.13}
\end{equation}
\begin{equation}
\langle
xy\rangle=\frac{\hbar\sin\phi\,\cos\phi}{2\sqrt{m_{x}m_{y}}}\left(\frac{1}
{\omega_{-}}-\frac{1}{\omega_{+}}\right).         \label{3.14}
\end{equation}
The oscillations of the ``neutron'' are described by $\langle
y^{2}\rangle$. In the most interesting case $k\gg q$ we have
\begin{equation}
\langle
y^{2}\rangle=\frac{\hbar}{2\sqrt{2mq}}\left(1+\sqrt{\frac{q}{4k}}
\right),                                     \label{3.15}
\end{equation}
In this approximation the result for the ``proton'' oscillations
$\langle x^{2}\rangle$ is the same. Again, the finite size of the
``deuteron'' leads to stretching of the zero-point oscillations by
a factor $(1+\sqrt{q/4k})$ and to the enhancement of the fusion
probability.

\subsection{Tunnelling of a bound electron (Cooper) pair}

The Hamiltonian of such a system may be modelled as
\begin{equation}
H=\frac{p_{x}^{2}+p_{y}^{2}}{2m}+\frac{1}{2}
k(x-y)^{2}+\frac{1}{2}q(x^{2}+y^{2}).               \label{3.16}
\end{equation}
Then the frequencies of two normal modes are
\begin{equation}
\omega_{+}^{2}=\frac{2k+q}{m}, \quad \omega_{-}^{2}=\frac{q}{m}.
                                              \label{3.17}
\end{equation}
At $q<0$, $\omega_{-}$ is imaginary. It does not depend on $k$,
therefore the parabolic barrier for center-of-mass motion is not
modified by the finite size of the system. However, there is the
energy transfer from the internal motion to the translational
motion if the potential barrier smoothly goes to zero at large
distances being parabolic near the top:
\begin{equation}
E=E(\infty)+\frac{\hbar\sqrt{2k/m}}{2}-\frac{\hbar\sqrt{(2k+q)/m}}{2}.
                                              \label{3.18}
\end{equation}
This energy transfer increases the tunnelling probability in the
case of a smooth barrier, e.g. at the the point contact. The case
of the tunnelling through the Josephson barrier is more
complicated since the adiabatic approximation is not valid there.
The situation might be closer to the regime of a sharp
perturbation that deserves a special consideration.

\section{Conclusion}

Our simple model helps to understand that, even in a two-particle
system, the tunnelling is a complex process that can be noticeably
influenced by the finite size of a tunnelling object and by its
intrinsic degrees of freedom. We pointed out the necessity to
study the behavior of the normal modes of a tunnelling system and
its adiabatic polarizability (or deformability) by the barrier
potential. The role of restructuring of complex particles in the
process of a nuclear reaction was emphasized in a different
context in refs. \cite{sakharuk97,sakharuk99}.

V.Z. acknowledges the useful discussions of different aspects of
the tunnelling problem for complex objects with C. Bertulani, M.
Horoi, M. Moshinsky, A. Sakharuk, and A. Volya, and the
discussions of the deuteron polarizability with W. Lynch. The work
was supported by the NSF, grants PHY-0070911 and PHY-0244453, and
by the Australian Research Council.

\end{document}